\documentclass[letterpaper,prb,twocolumn,showpacs]{revtex4}

\usepackage{graphicx,bm}
\usepackage{amsmath}
\usepackage{amsbsy}
\usepackage{amssymb}
\usepackage{array}
\usepackage{wasysym}
\newcommand{\sfrac}[2]{{\textstyle{\frac{#1}{#2}}}}

\newcommand{\bs}[1]{\boldsymbol{#1}}

\begin{document}
\title{Generic short-range interactions in two-leg ladders}
\author{J. E. Bunder}
\affiliation{Department of Physics, National Tsing-Hua University,
Hsinchu 300, Taiwan} \affiliation{Physics Division, National Center
for Theoretical Sciences, Hsinchu 300, Taiwan}
\author{Hsiu-Hau Lin}
\affiliation{Department of Physics, National Tsing-Hua University,
Hsinchu 300, Taiwan} \affiliation{Physics Division, National Center
for Theoretical Sciences, Hsinchu 300, Taiwan}

\begin{abstract}
We derive a Hamiltonian for a two-leg ladder which includes an
arbitrary number of charge and spin interactions. To illustrate this
Hamiltonian we consider two examples and use a renormalization group
technique to evaluate the ground state phases. The first example is a
two-leg ladder with zigzagged legs. We find that increasing the
number of interactions in such a two-leg ladder may result in a
richer phase diagram, particularly at half-filling where a few exotic
phases are possible when the number of interactions are large and the
angle of the zigzag is small. In the second example we determine
under which conditions a two-leg ladder at quarter-filling is able to
support a Tomanaga-Luttinger liquid phase. We show that this is only
possible when the spin interactions across the rungs are
ferromagnetic. In both examples we focus on lithium purple bronze, a
two-leg ladder with zigzagged legs which is though to support a
Tomanaga-Luttinger liquid phase.
\end{abstract}


\maketitle
\section{Introduction}

Ladder systems are well known for their many novel properties and
their relative simplicity makes them an ideal candidate for much
theoretical work.~\cite{Dagotto96,Maekawa96} Several experimental
systems are known to be of or dominated by a ladder-type structure,
and theoretical studies have been able to make reasonable predictions
about the phases, symmetries and transport properties of these
materials.~\cite{Scalapino95,Mayaffre98,Blumberg02} A common
procedure used to solve ladder systems is a perturbative
renormalization group (RG) treatment, followed by a non-perturbative
bosonization of the relevant interactions. The combination of these
two complimentary techniques allows one to go beyond the usual
mean-field approaches when determining ground states and excitations
in the low-energy regime.~\cite{Lin98} Some studies using these
techniques have revealed exotic phases, such as a staggered-flux
phase~\cite{Fjaerestad02} and a resonant-valence-bond
liquid.~\cite{Lin98}

\begin{figure}
\begin{center}
\includegraphics[width=6cm]{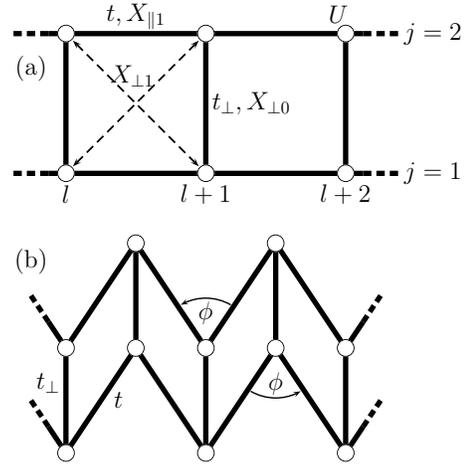}
\caption{(a) A standard two-leg ladder with hopping strengths $t$ and
$t_{\perp}$ and several electron-electron interactions defined
by $U$ and $X$ where $X=V,J$. (b) A zigzag two-leg ladder
where the legs are bent to make a constant angle $\phi$. Electron-electron
interactions can be defined similarly to the standard case.}
\label{fig:lattice}
\end{center}
\end{figure}

In a recent experiment,~\cite{Wang06} it was demonstrated that
Tomanaga-Luttinger liquid (TLL) behavior appears in lithium purple
bronze Li$_{0.9}$Mo$_6$O$_{17}$ (LPB). It is rather remarkable that
typical TLL scaling appears to exist over a wide range of
temperatures. In this reference it was claimed that RG flows
quantitatively reproduce the experimental data, but the bare
interactions which lead to this solution were not discussed. This
exciting development inspired us to revisit the well-known two-leg
ladder system, modified to describe a realistic interaction profile
while also taking into account the ladder geometry, as shown in Fig.
\ref{fig:lattice}.

A standard two-leg ladder is shown in Fig. \ref{fig:lattice}(a). The
hopping strengths between nearest neighbors on the same leg and
nearest neighbors on the same rung are $t$ and $t_{\perp}$
respectively. For on-site interactions the charge and spin
interactions take the same form and can be described by a single
parameter $U$. In general, interactions between two different lattice
sites along the same leg are described by $X_{\parallel n}$, while
interactions between two lattice sites on opposite legs are described
by $X_{\perp n}$. The charge and spin interactions are represented by
$X = V, J$ respectively and the integer index $n$ describes the rung
difference between the two sites. Therefore, any set of generic
short-range or quasi-long-range interactions can be described by the
bare interactions $U$, $X_{\parallel n}$ and $X_{\perp n}$. Although
there have been extensive theoretical investigations on electronic
correlations in two-leg
ladders,~\cite{Lin98,Fjaerestad02,Tsuchiizu02,Wu03,Tsuchiizu05,Chitov08}
most of these studies only consider nearest-neighbor (or
next-nearest-neighbor) interactions.

The standard two-leg ladder lies in a two-dimensional plane, but
there are a number of experimental systems which contain a two-leg
ladder which is warped in some fashion.~\cite{Popovic06,Brunger08} We
consider a ladder which has been compressed so that the legs form a
zigzag with a constant angle $\phi$, as shown in Fig.
\ref{fig:lattice}(b). One example of such a lattice is LPB which has
$\phi\sim \pi/2$.~\cite{Popovic06} By including the geometric
structure of the two-leg ladder we have an additional variable $\phi$
with which to investigate the ground-state phase diagram. It is easy
to see that for extremely short-range interactions the zigzag angle
$\phi$ does not play any significant role since the on-site
interaction $U$ dominates. However, in ladder materials the
interaction is often expected to be quasi-long-ranged and in these
cases $\phi$ is important. With the inclusion of this zigzag lattice
geometry, as well as the generic interactions we hope to not only be
able to fully describe a TLL phase but to also discover other exotic
phases such as an $f$-density wave and a staggered-flux phase.

The Tomanaga-Luttinger liquid is a special case amongst all the
possible phases of a two-leg ladder. In sharp contrast to ordinary
Fermi liquids, electron-electron interactions in TLL cause the
single-particle excitations (the so-called quasi-particles) to become
unstable. Instead one finds bosonic spin and charge excitations which
propagate independently of each other and with different velocities,
a phenomenon known as spin-charge separation. Many theoretical and
experimental studies have discussed TLL phase in several different 1D
or quasi-1D systems such as weakly-coupled chains or
wires,~\cite{Kim96,Segovia99} carbon
nanotubes,~\cite{Bockrath99,Ishii03} and the edges of two-dimensional
systems.~\cite{Wen90,Hilke01} With screened charge interactions
theoretical studies have shown that TLL are generally expected in
odd-leg ladders~\cite{Ledermann00} but not in even-leg ladders except
under unphysical conditions such as attractive
interactions.~\cite{Lin98} These general trends make the TLL scaling
behavior observed in LPB~\cite{Popovic06,Wang06} a little unexpected,
though as the two legs in LPB are almost independent ($t\gg
t_{\perp}$) it is certainly not impossible. There are two plausible
scenarios for the observed TLL-like behavior in LPB. The first
scenario is that the ground state is a true TLL and that the
interaction profile and the ladder geometry in LPB result in an
unusual set of bare couplings that flow towards the TLL phase under
RG transformations. The alternative possibility is that the ground
state is not a TLL but closely resembles a TLL over a wide range of
temperatures. In an attempt to solve this puzzle we will use LPB as
an example when determining the phases of our two-leg ladder with
zigzagged legs and generic interactions.

This paper is organized in the following way: In Sec. II, we
introduce the two-leg model that contains various charge and spin
interactions. Starting from the lattice model, we briefly describe
the chiral decomposition, current algebra, computation of initial
couplings and the bosonization. In Sec. III, we generalize the
theoretical approach to the two-leg ladder with general zigzag
angles. We also introduce different order parameters to characterize
the ground states. The complimentary combination of the RG method and
the bosonization techniques allows us to obtain the phase diagrams
for different interaction profiles and bending angles. We consider
two examples, the half-filled case with $t=t_{\perp}$ and the
quarter-filled case with $t\gg t_{\perp}$, with the latter case
corresponding to LPB. In Sec. IV, we make use of the general
theoretical framework developed in previous sections and try to
determine an appropriate interaction profile for a TLL in LPB. We
perform detailed and extensive numerical analysis and compute the
temperature-dependent TLL exponent. Finally, we conclude our
numerical results and discuss their connections to experiments.

\section{Model}

We consider a two-leg ladder with quasi-long-range charge and spin
interactions. The Hamiltonian contains non-interacting hopping as
well as charge and spin interactions over different ranges. Thus, it
is natural to divide the Hamiltonian into six parts,
\begin{equation}
H=H_0+H_U+H_{V_{\perp}}+H_{V_{\parallel}}
+H_{J_{\perp}}+H_{J_{\parallel}}.
\end{equation}
The first term $H_0$ describes hopping along the legs of the ladder
with hopping strength $t$, and along the rungs with hopping strength
$t_{\perp}$,
\begin{align}
H_0=&-t\sum_{jl\sigma}(c_{jl\sigma}^{\dag} c_{j(l+1)\sigma} +h.c.)\nonumber\\
&-t_{\perp}\sum_{l\sigma}(c_{1l\sigma}^{\dag}c_{2l\sigma} +h.c.).
\end{align}
The subscript of the fermion operator $c_{jl\sigma}$ describes leg
number $j=1,2$, rung number $l$, and spin
$\sigma=\uparrow,\downarrow$.

For the on-site interaction, the difference between the charge and
the spin parts vanishes and
\begin{equation}
H_U=U\sum_{jl}n_{jl\uparrow}n_{jl\downarrow}
\end{equation}
where $n_{jl\sigma}=c_{jl\sigma}^{\dag}c_{jl\sigma}$ and $U$ is the
interaction strength. Now we classify the more general charge
interactions.  Many theoretical studies consider perpendicular
nearest neighbor interactions across single rungs, i.e., between
sites $(j,l)$ and $(\bar{\jmath},l)$ where $\bar{\jmath}$ denotes the
opposite leg of $j$, as well as parallel nearest neighbor
interactions between neighboring sites on the same leg, i.e., between
sites $(j,l)$ and $(j,l\pm 1)$. A few studies also consider
next-nearest neighbor interactions which act diagonally across one
plaquette, i.e, between sites $(j,l)$ and $(\bar{\jmath},l\pm 1)$.
Here we consider all charge interactions between sites $(j,l)$ and
$(j',l')$ for which $|l-l'|\leq N$. Note that we have introduced a
``hard" cutoff length $N$ for the interaction profile. The
perpendicular Hamiltonian describes interactions between sites on
different legs
\begin{equation}
H_{V_{\perp}}=\sum_{n=0}^{N}
\sum_{jl\sigma\sigma'}V_{\perp n}
n_{jl\sigma}n_{\bar{\jmath} (l+n)\sigma'}
\end{equation}
where $V_{\perp n}$ is the interaction strength between sites $(j,l)$
and $(\bar{\jmath},l+n)$. The parallel Hamiltonian describes
interactions between sites on the same leg
\begin{equation}
H_{V_{\parallel}}=\sum_{n=1}^{N}
\sum_{jl\sigma\sigma'}V_{\parallel n}n_{jl\sigma}n_{j(l+n)\sigma'}
\end{equation}
where $V_{\parallel n}$ is the interaction strength between sites
$(j,l)$ and $(j,l+n)$.

Following the same classification the spin interactions are contained
in two parts, $H_{J_{\perp}}$ and $H_{J_{\parallel}}$. Like
$H_{V_{\perp}}$ and $H_{V_{\parallel}}$, the spin interaction
Hamiltonians describe interactions between any two sites which are
$N$ or less rung positions distant from each other. For spin
interactions between sites on different rungs,
\begin{equation}
H_{J_{\perp}}=\sum_{n=0}^{N}
\sum_{jl}J_{\perp n}
{\bf S}_{jl} \cdot {\bf S}_{\bar{\jmath} (l+n)}
\end{equation}
where $J_{\perp n}$ is the interaction strength between sites $(j,l)$
and $(\bar{\jmath},l+n)$ and the spin operators are
\begin{equation}
{\bf S}_{jl}=\sfrac{1}{2}\sum_{\sigma\sigma'} c^{\dag}_{jl\sigma}
\bm{\tau}_{\sigma\sigma'} c_{jl\sigma'},
\end{equation}
where $\bs{\tau}=(\tau_x,\tau_y,\tau_z)$ are Pauli matrices. For spin
interaction between different sites on the same leg,
\begin{equation}
H_{J_{\parallel}}=\sum_{n=1}^{N}
\sum_{jl\sigma\sigma'}J_{\parallel n}{\bf S}_{jl} \cdot
{\bf S}_{j(l+n)}
\end{equation}
where $J_{\parallel n}$ is the interaction strength between sites
$(j,l)$ and $(j,l+n)$.

We now follow a standard procedure which involves decomposing the
lattice fermions into pairs of chiral fermions with linear
dispersion. As this procedure is well explained
elsewhere~\cite{Lin98} we will only give a brief explanation.
Firstly, the hopping part of the Hamiltonian $H_0$ is diagonalized
into a bonding and antibonding band,
$a_{ql\sigma}=[c_{2l\sigma}-(-1)^qc_{1l\sigma}]/\sqrt{2}$ with
$q=1,2$, then after a Fourier transform we can determine the band
structure $E_q=(-1)^qt_{\perp}-2t\cos k_q$ as a function of momentum
$k_q$. The Fermi momentum
$k_{Fq}=\cos^{-1}[(-\mu+(-1)^qt_{\perp})/2t]$ is uniquely determined
by the chemical potential $\mu$. As we are only interested in the
low-energy behavior the fermion operators which diagonalize the
hopping Hamiltonian can be linearized about the Fermi point by
introducing chiral fermion fields, $a_{ql\sigma}\sim
\psi_{Rql\sigma}e^{ik_{Fq}l}+\psi_{Lql\sigma}e^{-ik_{Fq}l}$. Taking
the continuous limit of the discrete lattice index $l$, we can define
the Hamiltonian density $\mathcal{H}$ from $H=\int dl\mathcal{H}$.
The hopping part of the Hamiltonian density in terms of the chiral
fields is rather simple,
\begin{equation}
\mathcal{H}_0=-\sum_{q\sigma}v_q(\psi^{\dag}_{Rql\sigma}\partial_l
\psi_{Rql\sigma}-\psi^{\dag}_{Lql\sigma}\partial_l \psi_{Lql\sigma})
\end{equation}
where the Fermi velocity is $v_q=dE_q/dk_q$ at $k_q=k_{Fq}$.

The interaction part of the Hamiltonian density
$\mathcal{H}_I=\mathcal{H}_U+\mathcal{H}_{V_{\perp}}
+\mathcal{H}_{V_{\parallel}}+\mathcal{H}_{J_{\perp}}
+\mathcal{H}_{J_{\parallel}}$ can be expressed in terms of the
currents
\begin{alignat}{2}
J_{Pqq'}=&\sfrac{1}{2}\psi^{\dag}_{Pq\sigma} \psi_{Pq'\sigma},\,\,
&{\mathbf J}_{Pqq'}&=\sfrac{1}{2}\psi^{\dag}_{Pq\sigma}
\bs{\tau}_{\sigma\sigma'}\psi_{Pq'\sigma'}\nonumber\\
I_{Pqq'}=&\sfrac{1}{2}\psi_{Pq\sigma} \epsilon_{\sigma\sigma'}
\psi_{Pq'\sigma'},\,\, &{\mathbf
I}_{Pqq'}&=\sfrac{1}{2}\psi_{Pq\sigma}
(\epsilon\bs{\tau})_{\sigma\sigma'}\psi_{Pq'\sigma'}
\end{alignat}
where $P=R,L$. The antisymmetric matrix $\epsilon$ is defined by
$\epsilon_{12} = -\epsilon_{21}=1$ and $\epsilon_{11} =
\epsilon_{22}=0$. Each term in $\mathcal{H}_I$ is a product of two
currents so that the Hamiltonian is a function of four-fermion
interactions,
\begin{align}
\mathcal{H}_I=&b^{\rho}_{qq'}J_{Rqq'}J_{Lqq'}-b^{\sigma}_{qq'} {\mathbf J}_{Rqq'}.{\mathbf J}_{Lqq'}\nonumber\\
&+f^{\rho}_{qq'}J_{Rqq}J_{Lq'q'}-f^{\sigma}_{qq'}{\mathbf J}_{Rqq}.{\mathbf J}_{Lq'q'}\nonumber\\
&+u^{\rho}_{qq'}I^{\dag}_{qq'}I_{L\bar{q}\bar{q}'}
-u^{\sigma}_{qq'}{\mathbf I}_{Rqq'}.{\mathbf I}_{L\bar{q}\bar{q}'}.
\end{align}
The couplings of the four-fermion interactions $b_{qq'}$, $f_{qq'}$
and $u_{qq'}$ define the scattering amplitudes between bands $q$ and
$q'$. Backward scattering is represented by $b$ and from a gradient
expansion the bare coupling strength can be shown to be
\begin{align}
b^{\rho}_{qq}=&U+V_{\perp0}+2\sum_{n=1}^N[( V_{\parallel n}+V_{\perp n})(2-\cos 2nk_{Fq})]\nonumber\\
&-\sfrac{3}{4}J_{\perp0}-\sfrac{3}{2}\sum_{n=1}^N[(J_{\parallel n}+J_{\perp n})\cos 2nk_{Fq}]\nonumber\\
b^{\sigma}_{qq}=&U+V_{\perp0}+2\sum_{n=1}^N[( V_{\parallel n}+V_{\perp n})\cos 2nk_{Fq}]\nonumber\\
&-\sfrac{3}{4}J_{\perp0}-\sum_{n=1}^N[(J_{\parallel n}+J_{\perp n})(1+\sfrac{1}{2}\cos 2nk_{Fq})]\nonumber\\
b^{\rho}_{12}=&U-V_{\perp0}+2\sum_{n=1}^N[(V_{\parallel n}-V_{\perp n})(2\cos nk_--\cos nk_+)]\nonumber\\
&+\sfrac{3}{4}J_{\perp0}-\sfrac{3}{2}\sum_{n=1}^N[(J_{\parallel n}-J_{\perp n})\cos nk_+]\nonumber\\
b^{\sigma}_{12}=&U-V_{\perp0}+2\sum_{n=1}^N[(V_{\parallel n}-V_{\perp n})\cos nk_+]\nonumber\\
&+\sfrac{3}{4}J_{\perp0}-\sum_{n=1}^N[(J_{\parallel n}-J_{\perp n})(\cos nk_-+\sfrac{1}{2}\cos nk_+)]\label{eq:b}
\end{align}
where $k_{\pm}=k_{F1}\pm k_{F2}$. The symmetry of the system requires
$b_{12}=b_{21}$ and at half-filling $\mu=0$ so $k_{F1}+k_{F2}=\pi$
which sets $b_{11}=b_{22}$. Forward scattering is represented by $f$
with the bare coupling strength,
\begin{align}
f^{\rho}_{12}=&U+2\sum_{n=1}^N[V_{\parallel n}(2-\cos nk_+)+V_{\perp n}(2+\cos nk_+)]\nonumber\\
&+3V_{\perp0}+\sfrac{3}{4}J_{\perp0}
-\sfrac{3}{2}\sum_{n=1}^N[(J_{\parallel n}-J_{\perp n})\cos nk_+]\nonumber\\
f^{\sigma}_{12}=&U-V_{\perp0}+2\sum_{n=1}^N[(V_{\parallel n}-V_{\perp n})\cos nk_+]-\sfrac{1}{4}J_{\perp0}\nonumber\\
&-\sum_{n=1}^N[J_{\parallel n}(1+\sfrac{1}{2}\cos nk_+)-J_{\perp n}(1-\sfrac{1}{2}\cos nk_+)].\label{eq:f}
\end{align}
Symmetry requires $f_{12}=f_{21}$ and in order to avoid double
counting we set $f_{qq}=0$. Umklapp scattering is represented by $u$
and only present at half-filling where $k_{F1} + k_{F2} = \pi$,
\begin{align}
u^{\rho}_{11}=&U-V_{\perp0}+2\sum_{n=1}^N[(V_{\parallel n}-V_{\perp n})(-1)^n]\nonumber\\
&+\sfrac{3}{4}J_{\perp0}-\sfrac{3}{2}\sum_{n=1}^N
[(J_{\parallel n}-J_{\perp n})(-1)^n]\nonumber\\
u^{\rho}_{12}=&2U+2\sum_{n=1}^N[V_{\parallel n}(\cos 2nk_{F1}+(-1)^n)\nonumber\\
&+V_{\perp n}(\cos 2nk_{F1}-(-1)^n)]\nonumber\\
&-\sfrac{3}{2}\sum_{n=1}^N[J_{\parallel n}(\cos 2nk_{F1}+(-1)^n)
\nonumber\\
&-J_{\perp n}(\cos 2nk_{F1}-(-1)^n)]\nonumber\\
u^{\sigma}_{12}=&2V_{\perp0}
+2\sum_{n=1}^N[V_{\parallel n}(\cos 2nk_{F1}-(-1)^n)\nonumber\\
&+V_{\perp n}(\cos 2nk_{F1}+(-1)^n)+\sfrac{1}{2}J_{\perp0}\nonumber\\
&+\sfrac{1}{2}\sum_{n=1}^N[J_{\parallel n}(\cos 2nk_{F1}-(-1)^n)\nonumber\\
&+J_{\perp n}(\cos 2nk_{F1}+(-1)^n)]\label{eq:u}
\end{align}
with $u_{12}=u_{21}$ and $u_{11}=u_{22}$ from symmetry. At
half-filling, the particle-hole symmetry ensures we have nine unique
coupling constants, $b^{\rho}_{11}=b^{\rho}_{22}$,
$b^{\sigma}_{11}=b^{\sigma}_{22}$, $b^{\rho}_{12}$,
$b^{\sigma}_{12}$, $f^{\rho}_{12}$, $f^{\sigma}_{12}$,
$u^{\rho}_{11}=u^{\rho}_{22}$, $u^{\rho}_{12}$ and $u^{\sigma}_{12}$.
Away from half-filling the Umklapp interactions vanish but we no
longer have $b_{11}=b_{22}$ so we have eight different coupling
constants.

The Hamiltonian is more easily analyzed if the chiral fermion
operators are replaced with boson operators.~\cite{Lin98,vonDelft98}
The bosonized fields $\theta_{\nu\pm}$ and $\varphi_{\nu\pm}$ with
$\nu=\rho,\sigma$ represent a variety of quantum numbers. The
subscript represents total (+) or relative (-) charges or spins
($\rho$ or $\sigma$ respectively) between the two bands while
$\theta$ is a displacement field and $\varphi$ is a phase field. The
total bosonized Hamiltonian density
$\mathcal{H}=\mathcal{H}_0+\mathcal{H}_I$ is
\begin{multline}
\mathcal{H}=\frac{1}{8\pi}\sum_{\nu\pm}v_{\nu\pm}[ K_{\nu\pm}^{-1}
(\partial_x \theta_{\nu\pm})^2+K_{\nu\pm}(\partial_x\varphi_{\nu\pm})^2]\\
-2b^{\sigma}_{12}\cos\varphi_{\rho-}\cos\theta_{\sigma+}
+2\cos\theta_{\sigma+}(b^{\sigma}_{11}\cos\theta_{\sigma-}\\
+f^{\sigma}_{12}\cos\varphi_{\sigma-})-\cos\varphi_{\rho-}
(b^+_{12}\cos\theta_{\sigma-}+b^-_{12}\cos\varphi_{\sigma-})\\
-2u^{\rho}_{11}\cos\theta_{\rho+}\cos\varphi_{\rho-}
-2u^{\sigma}_{12}\cos\theta_{\rho+}\cos\theta_{\sigma+}\\
-\cos\theta_{\rho+}(u^+_{12}\cos\theta_{\sigma-} +
u^-_{12}\cos\varphi_{\sigma-}) \label{eq:bosonized}
\end{multline}
where $b^{\pm}_{12}=b^{\sigma}_{12}\pm b^{\rho}_{12}$ and
$u^{\pm}_{12}=u^{\sigma}_{12}\pm u^{\rho}_{12}$. The Luttinger
parameters and the Fermi velocities for the total/relative charge and
spin sectors are
\begin{align}
K_{\nu\pm}=&\sqrt{\frac{2\pi(v_1+v_2)-
[(b^{\nu}_{11}+b^{\nu}_{22})/2\pm f^{\nu}_{12}]}{2\pi(v_1+v_2)
+[(b^{\nu}_{11}+b^{\nu}_{22})/2\pm f^{\nu}_{12}]}}\label{eq:Krho}\\
v_{\nu\pm}=&
\sqrt{4\pi^2(v_1+v_2)^2-[(b^{\nu}_{11}+b^{\nu}_{22})/2\pm f^{\nu}_{12}]^2}/4\pi.\nonumber
\end{align}
Note that the highly symmetric bosonized form in Eq.
(\ref{eq:bosonized}) is possible only for degenerate velocity $v_1 =
v_2$. This is always true at half-filling but not at generic
fillings. The LPB two-leg ladder we are interested in is at
quarter-filling,~\cite{Popovic06} but as it consists of nearly
independent chains with vanishingly small inter-chain hopping
$t_{\perp} \ll t$, the Fermi velocities are nearly degenerate
$v_1\sim v_2$ and the above bosonized Hamiltonian is valid.

The RG flow equations of the couplings are of the form $d g_{i}/
d\ell =\sum_{jk} A^{jk}_{i}g_{j}g_{k}$, where $A^{jk}_i$ is a
constant tensor that can be computed from operator product
expansions.~\cite{Lin98,Chang05} All RG flow equations are solved
simultaneously with the initial conditions at $\ell=0$ given in Eqs.
(\ref{eq:b}-\ref{eq:u}). The ground state phase is determined from a
non-zero order parameter, such as electron density or current flow,
which can be evaluated using the solutions of the RG flow equations.
On substituting the RG solutions into the bosonized Hamiltonian Eq.
(\ref{eq:bosonized}) the Hamiltonian may be minimized by a specific
set of pinned bosonized fields (while other fields remain free to
adopt any value), thus defining the ground state. For example, if
$b^{\sigma}_{12}$ flows to a non-zero value then
$b^{\sigma}_{12}\cos\varphi_{\rho-}\cos\theta_{\sigma+}$ in Eq.
(\ref{eq:bosonized}) could minimize the Hamiltonian by pinning
$\varphi_{\rho-},\theta_{\sigma+}=m\pi$ for some integer $m$. In
order to maintain this minimum, $m$ can change by integral values,
which describes an excitation over some finite energy gap. If instead
$b^{\sigma}_{12}$ flows to zero the term
$b^{\sigma}_{12}\cos\varphi_{\rho-}\cos\theta_{\sigma+}$ is
irrelevant. Any field that remains unpinned when the Hamiltonian is
minimized may describe a gapless excitation. The pinned fields may be
substituted into the order parameter equations, once they are
appropriately bosonized, to determine the phase. Therefore, each
phase can essentially be defined by a unique set of pinned boson
fields, which are directly related to the solutions of the RG flow
equations. Note that it is only the coefficients of the sinusoidal
terms which ultimately determine the gapped excitations and therefore
$b^{\rho}_{qq}$ and $f^{\rho}_{12}$ are the only couplings which can
be non-zero in a fully gapless phase, i.e., a TLL.

\section{Zigzag two-leg ladder}
Now we would like to incorporate the realistic ladder geometry into
the above theoretical model. We consider a two-leg ladder in which
the legs zigzag parallel to each other, as shown in Fig.
\ref{fig:lattice}(b). The bent legs make a constant angle $\phi$. For
unscreened charge interactions the interaction strengths between two
sites are inversely proportional to the distance between them so,
\begin{align}
X_{\parallel n}=&\frac{X}{an\sin\phi/2},\qquad n = 2,4,6,...
\nonumber\\
X_{\parallel n}=&\frac{X}{a\sqrt{1+(n^2-1)\sin\phi/2}},\qquad n=1,3,5,...
\nonumber\\
X_{\perp n}=&\frac{X}{a\sqrt{\delta^2+n^2\sin^2\phi/2}},\qquad n=0,2,4,...
\nonumber\\
X_{\perp n}=&\frac{X}{a\sqrt{1+\delta^2+(n^2-1)\sin\phi/2}},\qquad n=1,3,5,...
\end{align}
where $X=V,J$, $a$ is the distance between neighboring lattice sites
on the same leg and the distance between lattice sites on the same
rung is $a\delta$. We shall assume the ladder consists of square
plaquettes with $\delta =1$. Generally we would like interactions
beyond the cutoff, i.e., with $n>N$, to be less strong than
interactions within the cutoff, but when $\phi$ is very small this
may not be the case. This issue may be avoided by defining different
cutoffs for interactions along a leg and interactions between legs,
but as the small values of $\phi$ for which this problem occurs are
quite likely not experimentally attainable we will continue to use
just one cutoff $N$.

\subsection{Order parameters}

At half-filling the phase of a two-leg ladder could be one of four
density wave phases or one of four Mott insulator
phases.~\cite{Tsuchiizu02,Tsuchiizu05} We first discuss the density
wave phases, the charge density wave (CDW), the staggered-flux (SF)
phase, the $p$-density wave (PDW) and the $f$-density wave (FDW), and
their associated order parameters. A CDW has a non-zero variation in
the average electron density per site which is defined by
\begin{equation}
n_{jl}=\sum_{\sigma}c^{\dag}_{jl\sigma}c_{jl\sigma}.
\end{equation}
At half-filling the average electron density of a two-leg ladder is
one electron per site, but in a CDW the sites are alternatively
unoccupied or fully occupied by two electrons. To define current flow
we use
\begin{align}
j_{\perp jl}=&i\sum_{\sigma}[c^{\dag}_{\bar{\jmath}l\sigma}c_{jl\sigma} -h.c.]\nonumber\\
j_{\parallel jl}=&i\sum_{\sigma}[c^{\dag}_{j(l+1)\sigma}c_{jl\sigma} -h.c.]\nonumber\\
j_{d jl}=&i\sum_{\sigma}[c^{\dag}_{\bar{\jmath}(l+1) \sigma}c_{jl\sigma} -h.c.]
\end{align}
which describe currents along rungs between sites $(j,l)$ and
$(\bar{\jmath},l)$, along legs between sites $(j,l)$ and $(j,l+1)$
and along the diagonals of the plaquettes between sites $(j,l)$ and
$(\bar{\jmath},l+1)$, respectively. If the first two currents are
non-zero we have a SF phase which is characterized by alternative
clockwise and anticlockwise current flows around plaquettes. If the
third current is non-zero we have a FDW which is characterized by
currents zigzagging across the diagonals of the plaquettes. Two
kinetic order parameters are
\begin{align}
B_{\parallel jl}=&i\sum_{\sigma}[c^{\dag}_{j(l+1)\sigma}c_{jl\sigma} +h.c.]\nonumber\\
B_{d jl}=&i\sum_{\sigma}[c^{\dag}_{\bar{\jmath}(l+1) \sigma}c_{jl\sigma} +h.c.]
\end{align}
where $B_{\parallel jl}$ describes interactions along legs and $B_{d
jl}$ describes interactions across diagonals. A PDW is defined by
non-zero $B_{\parallel jl}$ which implies dimerization between
neighboring sites on the same leg. While non-zero $B_{d jl}$ does not
formally define any phase, it tends to be non-zero in a CDW and
describes dimerization between sites of equal electron density. The
kinetic energy across rungs
\begin{equation}
B_{\perp jl}=i\sum_{\sigma}[c^{\dag}_{\bar{\jmath}l\sigma}c_{jl\sigma} +h.c.]
\end{equation}
is always zero.

Away from half-filling the situation is a little different with there
being only two possible density waves phases. In this case a CDW (SF)
and a PDW (FDW) coexist in a single phase, and for simplicity we name
this phase a CDW (SF) phase. At half-filling the CDW (SF) and the PDW
(FDW) only differ by the pinned value of the total charge
displacement field $\theta_{\rho+}$. Away from half-filling the
Umklapp terms are removed and this provides an additional symmetry,
resulting in an unpinned $\theta_{\rho+}$. When $\theta_{\rho+}$ is
unpinned the CDW and PDW may coexist with the relevant order
parameters, $n_{jl}$ and $B_{\parallel jl}$, being simultaneously
non-zero. Similarly, the SF and the FDW may coexist and all three
currents $j_{\perp jl}$, $j_{\parallel jl}$ and $j_{d jl}$ will be
simultaneously non-zero.

If all the order parameters discussed above vanish then we may have a
Mott insulator or a superconductor state. An $s$-wave superconducting
order parameter can be defined by
\begin{equation}
\Delta_{slj}=c_{jl\uparrow}c_{jl\downarrow}\sim\frac{1}{2}\sum_{Pq}
\Delta_{Pql}
\end{equation}
where $\Delta_{Pql}=\psi_{Pql\uparrow}\psi_{\bar{P}ql\downarrow}$ is
the pairing operator of the chiral fields.  The $d$-wave order
parameter across the rungs is
\begin{equation}
\Delta_{d\perp l}=c_{1l\uparrow}c_{2l\downarrow}\sim\frac{1}{2}\sum_{Pq}
(-1)^{q+1}\Delta_{Pql}.
\end{equation}
As the names imply, $\Delta_{slj}$ is non-zero in an $s$-wave
superconductor (S-SC) while $\Delta_{d\perp l}$ is non-zero in an
$d$-wave superconductor (D-SC). On bosonizing the superconducting
order parameters it can be seen that they can only be non-zero away
from half-filling where the boson field $\theta_{\rho+}$ is unpinned.
At half-filling the total charge displacement is pinned and both
$\Delta_{slj}$ and $\Delta_{d\perp l}$ vanish, and provided all
previously discussed order parameters are also zero we may have a
Mott insulator. The Mott insulator at half-filling is defined by
non-zero $\Delta_{Pql}$ and, like a superconductor, is defined in
terms of a pairing symmetry. If $\Delta_{R1l}\Delta^{\dag}_{R2l}>0$
we define the Mott insulator as $s$-wave, but if
$\Delta_{R1l}\Delta^{\dag}_{R2l}<0$ we define it as $d$-wave. Two
types of $s$-wave and $d$-wave Mott insulators exist, one with
$\theta_{\rho+}$ pinned to an even multiple of $\pi$, named S-Mott
and D-Mott, and the other with $\theta_{\rho+}$ pinned to an odd
multiple, named S$'$-Mott and D$'$-Mott. A difference in
$\theta_{\rho+}$ of $\pi$ represents a half-plaquette shift in the
center of mass of the paired chiral fields with the D and S-Mott
pairing being across rungs and the D$'$ and S$'$-Mott pairing being
across the diagonals of the plaquettes.

There are a few other possible phases in the two-leg ladder. For
example, the phases transitions between any two phases may be though
of as phases in their own right, but as they only exist over a
vanishingly small parameter range we will not discuss them here.
Finally we should mention the TLL, although we shall discuss this
phase in more detail in Sec. \ref{sec:LL}. In a TLL all bosonic
fields are unpinned and because of this all order parameters
discussed above are undefined.

\subsection{Half-filling}

\begin{figure}
\begin{center}
\includegraphics[width=8cm]{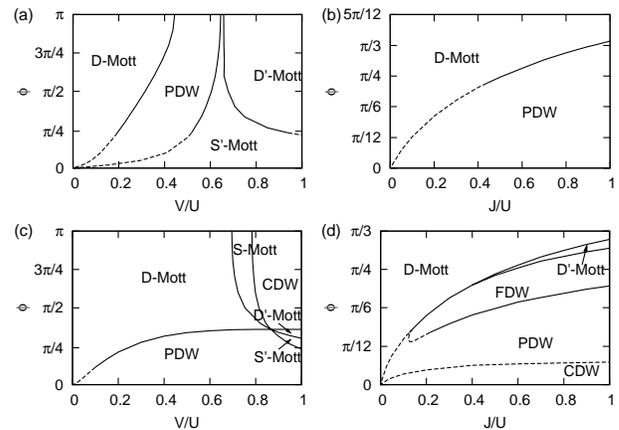}
\caption{Phase diagrams at half-filling with $t=t_{\perp}=1$, $U=2$ and (a) $N=2$, $J=0$ (b) $N=2$,
$V=0$ (c) $N=10$, $J=0$ (d) $N=10$, $V=0$.  Dashed lines indicate where
$\phi$ is small enough so that some interactions not considered are
larger than some which are considered. }
\label{fig:phase2-10}
\end{center}
\end{figure}

\begin{figure}
\begin{center}
\includegraphics[width=8cm]{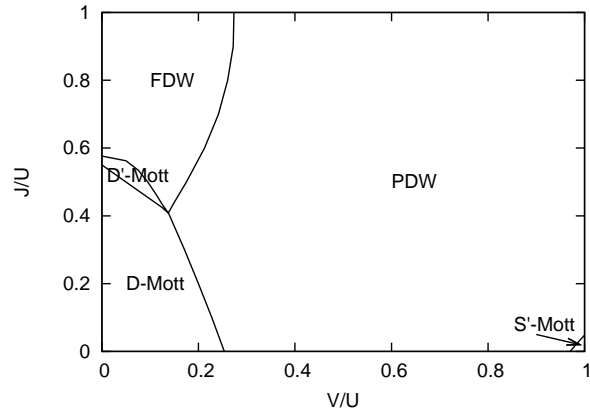}
\caption{Phase diagram at half-filling with $N=10$,
$t=t_{\perp}=1$, $U=2$ and $\phi=\pi/4$. }
\label{fig:phase10}
\end{center}
\end{figure}

Our first example of the zigzag two-leg ladder is the case of
half-filling $\mu=0$ with equal leg and rung hopping $t=t_{\perp}$.
Phase diagrams constructed from the solutions of the RG flow
equations are given in Fig. \ref{fig:phase2-10} and clearly both
$\phi$ and $N$ play a significant role. We always assume on-site
interaction $U=2$, although in the results presented here it is the
ratios $V/U$ and $J/U$ which are important in determining the phase,
rather than the actual values of $U$, $V$ and $J$. When there are
charge interactions $V\neq 0$ but no spin interactions $J=0$
increasing $N$ from 2 to 10 will allow CDW and S-Mott phases to
emerge while significantly reducing the range of the D$'$ and
S$'$-Mott states. For spin interactions $J\neq 0$ but no charge
interactions $V=0$, when $N=2$ only a D-Mott and a PDW phase are
possible, and only at quite small angles, $\phi\lesssim\pi/3$. As $N$
is increased to 10 one still requires $\phi\lesssim\pi/3$ to obtain
anything but a D-Mott phase, but at these small angles several phases
are possible. Of particular interest is the emergence of a FDW as
this phase has not previously been predicted in a two-leg ladder at
half-filling under any physically possible scenarios. Although the
angle required to obtain a FDW is quite small it may be possible to
construct an appropriate lattice using cold atoms.~\cite{Jaksch05}
The phase diagram for $N=10$ and $\phi=\pi/4$ with variable $V$ and
$J$ is shown in Fig. \ref{fig:phase10}. Although the PDW dominates,
there is quite a substantial FDW region.

\subsection{Quarter-filling}
\begin{figure}
\begin{center}
\includegraphics[width=8cm]{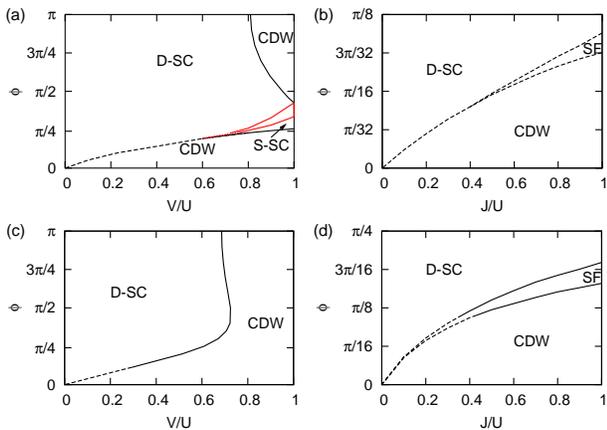}
\caption{(Color online) Phase diagrams at quarter-filling with $t=1$,
$t_{\perp}=0.01$, $U=2$ and (a) $N=2$, $J=0$ (b) $N=2$,
$V=0$ (c) $N=10$, $J=0$ (d) $N=10$, $V=0$.  Dashed lines indicate where
$\phi$ is small enough so that some interactions not considered are
larger than some which are considered. }
\label{fig:phaseLPB2-10}
\end{center}
\end{figure}

For our second example of the zigzag two-leg ladder we assume
quarter-filling which  sets $\mu=-\sqrt{2t^2-t_{\perp}^2}$ and we
also assume $t=1\gg t_{\perp}=0.01$. This case is designed to
correspond to LPB when $\phi\sim \pi/2$. We again use $U=2$,
although, as before, it is the ratios $V/U$ and $J/U$ which
ultimately determine the phase. In Fig. \ref{fig:phaseLPB2-10} we
present a number of phase diagrams. Despite the small SF phase, the
$J\neq 0$, $V=0$ case is not particularly interesting as extremely
small values of $\phi$ are required if any phase but a D-SC is to be
observed, particularly when $N=2$. Unlike the half-filled case,
increasing $N$ from 2 to 10 does not cause new phases to emerge,
although the SF phase does appear at a larger value of $\phi$. In the
$V\neq 0$, $J=0$ case increasing $N$ from 2 to 10 decreases the
complexity of the phase diagram, causing the CDW phase to expand and
the S-SC phase to vanish.

The region in Fig. \ref{fig:phaseLPB2-10}(a) which is marked in red
describes a region of unusual scaling. The phase in this region is
either D-SC or S-SC, with the D-SC to S-SC phase transition running
approximately through the center. Each phase is characterized by a
unique set of RG solutions of the eight coupling constants and
generally, while renormalizing, the coupling constants flow gradually
towards this final final solution. In the red region the coupling
constants do not initially flow towards either a D-SC or S-SC
solution but instead towards a solution typical of the D-SC to S-SC
phase transition. This scaling behavior continues as $\ell$
increases, but at some point there is a sudden change and the RG will
flow rapidly to either a D-SC or S-SC solution. This scaling behavior
is typical when extremely close to a phase transition, but it is not
usually observed in regions as large as the red region in Fig.
\ref{fig:phaseLPB2-10}(a). It is quite possible that this region
could be mistaken for a TLL phase, as we shall explain in more detail
in the next section. Note that this region is very close to
$\phi=\pi/2$ so it may explain the TLL observations in
LPB.~\cite{Wang06}

\section{Tomanaga-Luttinger liquid}
\label{sec:LL} In the previous section we constructed various phase
diagrams while assuming physically realistic conditions, yet did not
observe a TLL. In this section we look more closely at what is
required for the RG equations to flow towards a TLL solution. We
simplify the problem a little by considering a two-leg ladder with
the interaction cutoff $N=1$. Note that in this limit, the zigzag
angle $\phi$ does not affect the initial values of the couplings and
thus can be ignored. This model was discussed in Ref.
\onlinecite{Wang06} to describe LPB and solved using RG flow
equations equivalent to the ones used here. Note that one of the key
features for TLL is the critical exponent $\alpha$ of the
single-particle density of states. Quasi-particle excitations are not
found in a zero-temperature TLL so the single-particle density of
states $\rho(\epsilon)$ at energy $\epsilon$ should be suppressed
near the Fermi energy $\epsilon_F$. This suppression is expected to
follow a power law
$\rho(\epsilon)\propto|\epsilon-\epsilon_F|^{\alpha}$ for some
positive constant $\alpha$ as the temperature approaches
zero.~\cite{Voit94}

In Ref. \onlinecite{Wang06}, it was argued, both experimentally and
theoretically, that the nature of the critical exponent $\alpha$
indicates that LPB has a TLL phase. A remarkable agreement was found
between the experimental value of $\alpha$ and the theoretical value
obtained from the RG solutions, but what electron-electron
interactions would provided the required initial conditions of the RG
equations were not stated. Here we will discuss the electron-electron
interactions which may support a TLL in a LPB-like two-leg ladder.

We can evaluate $\alpha$ numerically at different temperatures from
the coupled RG equations. Note that the the temperature scales as
$T=T_{0}e^{-\ell}$ under RG transformations, where $T_0$ is the
initial (bare) temperature. Therefore, when calculating the
couplings' flow with the logarithmic length scale $\ell$, we can
compute the critical exponent $\alpha$ at different temperatures. It
is known that the exponent takes the form,
\begin{equation}
\alpha=(K_{\rho+}+K_{\rho+}^{-1}+
K_{\rho-}+K_{\rho-}^{-1}-4)/8. \label{eq:exponent}
\end{equation}
If it approaches a constant during the RG analysis we have a hint of
TLL behaviour.~\cite{Wang06} For convenience we separate this
critical exponent into two parts, $\alpha=\alpha_+ +\alpha_-$ where
\begin{equation}
\alpha_{\pm}=(K_{\rho\pm}+K_{\rho\pm}^{-1}-2)/8.
\end{equation}
As has been discussed previously, the couplings in front of the
sinusoidal terms in Eq. (\ref{eq:bosonized}) determine the energy
gaps and thus the nature of the phase. If none of these couplings
become relevant under RG transformation, the ground state is gapless
and is characterize by the so-called Luttinger parameters
$K_{\rho\pm}$ and $K_{\sigma\pm}$ in the charge and spin sectors. In
this case only the first line of Eq. (\ref{eq:bosonized}) remains,
which corresponds to the TLL Hamiltonian.~\cite{vonDelft98} If at
least one of the coefficients of the sinusoidal terms does not flow
to zero we have any one of the Mott, SC, or density wave states
discussed above. Three couplings, $b^{\rho}_{qq}$ with $q=1,2$ and
$f^{\rho}_{12}$, are not coefficients of sinusoidal terms so need not
vanish in a TLL. From the RG equations it can be seen that these
three couplings will remain roughly constant when, and only when, all
the other gap-inducing couplings are irrelevant.~\cite{Lin98,Chang05}
Furthermore, only these three couplings appear in Eq.
(\ref{eq:exponent}) which defines $\alpha$. This is in agreement with
what we have already stated about a TLL, i.e. the RG solution of
$\alpha$ must flow to a constant value.

We can make some comments about the general behavior of $\alpha$.
From the RG flow equations we determine that $b^{\rho}_{qq}$ always
decreases, but $f^{\rho}_{12}$ always increases. In fact,
$b^{\rho}_{11}+b^{\rho}_{22}+2f^{\rho}_{12}$ remains constant in RG
flows. Therefore $K_{\rho+}$ must be a constant and $K_{\rho-}$
always increases. This implies $\alpha_+$ is a constant and the flow
of the exponent $\alpha$ is essentially determined by $\alpha_-$. The
minimum of $\alpha$ will be when $\alpha=\alpha_+$, which corresponds
to $\alpha_-=0$ and $K_{\rho-}=1$, or equivalently $b^{\rho}_{11} +
b^{\rho}_{22}=2f^{\rho}_{12}$. When $K_{\rho-}<1$ (or $b^{\rho}_{11}
+ b^{\rho}_{22}>2f^{\rho}_{12}$) $\alpha_-$ will decrease, as must
$\alpha$, but when $K_{\rho-}>1$ both $\alpha_-$ and $\alpha$ will
increase. The point  $b^{\rho}_{11} + b^{\rho}_{22}=2f^{\rho}_{12}$
is more significant that just the turning point of $\alpha$. The RG
flow equations indicate that once this point has been reached our
system cannot be a TLL and $\alpha$ will increase at an increasing
rate. So, once $b^{\rho}_{11} + b^{\rho}_{22} < 2f^{\rho}_{12}$, or
equivalently $K_{\rho-}>1$, in the RG flows, the TLL phase is
unstable and some energy gaps will appear. However, it is important
to emphasize that with $b^{\rho}_{11} + b^{\rho}_{22} >
2f^{\rho}_{12}$ satisfied we may have a TLL but it is not guaranteed.

\begin{figure}
\begin{center}
\includegraphics[width=8cm]{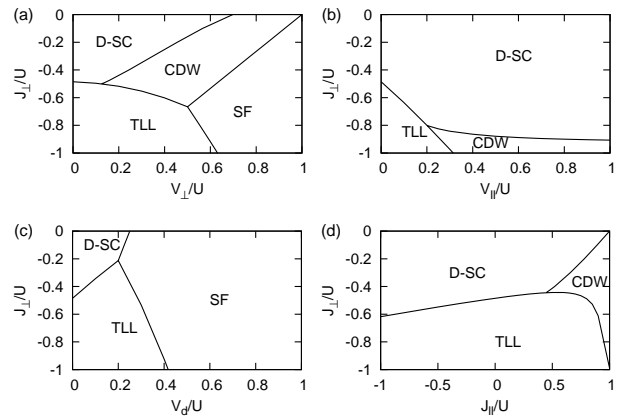}
\caption{Phase diagrams at quarter-filling with $t=1$, $t_{\perp}=0.01$,
$U=2$ and (a)
$V_{\parallel}=V_d=J_{\parallel}=0$, (b) $V_{\perp}=V_d=J_{\parallel}=0$,
 (c) $V_{\perp}=V_{\parallel}=J_{\parallel}=0$, (d)
 $V_{\perp}=V_{\parallel}=V_d=0$}
\label{fig:phase}
\end{center}
\end{figure}

Returning to our specific example of LPB, we again assume
quarter-filling and set $t=1$, $t_{\perp}=0.01$. In this case
$k_{F1},k_{F2}\sim\pi/4$. Using the initial conditions in Eqs.
(\ref{eq:b}) and (\ref{eq:f}) it can be shown that the condition
$b^{\rho}_{11}+b^{\rho}_{22}>2f^{\rho}_{12}$ is equivalent to
\begin{equation}
J_{\perp0}+2\sum_{n=1}^{N/2}(-1)^nJ_{\perp2n}<
-\sfrac{4}{3}[V_{\perp0}+2\sum_{n=1}^{N/2}(-1)^nV_{\perp2n}].
\end{equation}
In order to compare our results with Ref. \onlinecite{Wang06} we only
consider the charge interactions $V_{\perp}=V_{\perp0}$,
$V_{d}=V_{\perp1}$, $V_{\parallel}=V_{\parallel 1}$ and the spin
interactions $J_{\perp}=J_{\perp0}$, and $J_{\parallel}=J_{\parallel
1}$ which corresponds to a cutoff $N=1$, and so the necessary but not
sufficient condition for a TLL reduces to
$J_{\perp}<-\sfrac{4}{3}V_{\perp}$. We wish to restrict ourselves to
physically possible cases so we must have $V_{\perp}\geq 0$ and
therefore the spin interaction across rungs $J_{\perp}$ must be
negative, implying ferromagnetic exchange coupling. In Fig.
\ref{fig:phase} we present four phase diagrams, all of which show a
LL may be obtained for negative $J_{\perp}$. In all cases the
condition $J_{\perp}<-\sfrac{4}{3}V_{\perp}$ is satisfied when we
have a TLL, but clearly it does not imply that we must have a TLL.

\begin{figure}
\begin{center}
\includegraphics[width=8cm]{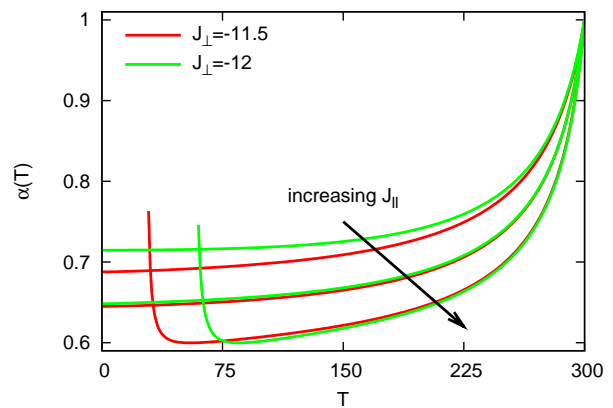}
\caption{(Color online) The critical exponent $\alpha$ at quarter-filling with
$t=1$, $t_{\perp}=0.01$, $U=2$,$V_{\parallel}=V_{\perp}$ and $J_{\parallel}=(1 + 2j)$
with $j=0,1,2$. While the arrow indicates increasing $J_{\parallel}$ for the
curves show here, it does not indicate a general trend. The lowest red
line and the lowest green line diverge so are not TLL, while the other
four lines approach constant values and imply a TLL state.}
\label{fig:purpbronze}
\end{center}
\end{figure}

If we wish to choose initial conditions which will enable the RG flow
of $\alpha$ to closely resemble the experimental data in Ref.
\onlinecite{Wang06} then even greater restrictions are placed on
$J_{\perp}$. As the interaction strengths are unknown we must attempt
to make a reasonable guess. We set $U=2$ and to make the numerical
search practical we set $V_{\perp} = V_{\parallel}$. Then, we fit the
experimental data by varying the bare values of the charge
interactions $V_{\perp}, V_{d}$ and the spin interactions $J_{\perp},
J_{\parallel}$. According to experimental data~\cite{Wang06}
$0.6\lesssim \alpha<1$, with $\alpha=1$ corresponding to the highest
temperature measurement. So, we set $\alpha_+=0.6$ as this marks the
minimum of $\alpha$ and this sets $K_{\rho+}=0.15$. The maximum value
is set to $\alpha(T_{0})=1$ so
$\alpha_-(T_0)=\alpha(T_{0})-\alpha_+=0.4$ and
$K_{\rho-}(\ell=0)=0.20$. These conditions determine the initial
values of $b^{\rho}_{11}+ b^{\rho}_{22}\pm 2f^{\rho}_{12}$ which in
turn determine $V_{\perp}, V_{d}$ for a given choice of $J_{\perp},
J_{\parallel}$. The RG flow of $\alpha$ which corresponds to the
experimental data is shown in Fig. \ref{fig:purpbronze}, with the
initial temperate $T_{0}=300$ K. We find that a TLL with $0.6\lesssim
\alpha<1$ is only obtained when $J_{\perp}$ is quite large
(significantly larger than $U$) and negative, which is unrealistic
for LPB. This does not mean that a TLL phase is impossible for LPB as
we must bear in mind that these simple one-loop RG solutions are not
expected to be quantitatively correct and should really only be used
for qualitative analysis. Consequently, forcing the theoretical value
of $\alpha$ to fit the experimental data is not recommended and will
not give a good prediction of the interactions in the lattice.

\begin{figure}
\begin{center}
\includegraphics[width=8cm]{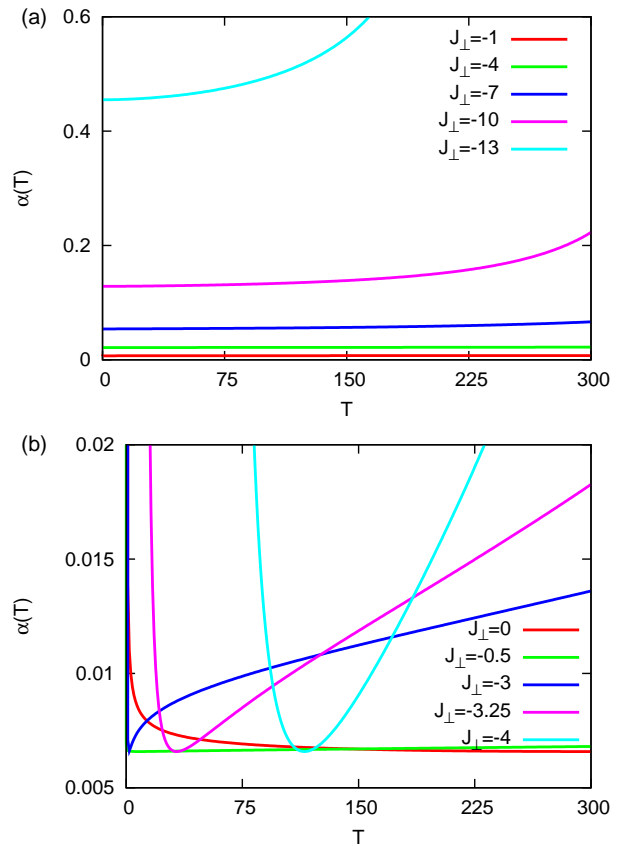}
\caption{(Color online) The critical exponent $\alpha$ at quarter-filling with
$t=1$, $t_{\perp}=0.01$, $U=2$ and (a) $J_{\parallel}=V_{\perp}=
V_{\parallel}=V_d=0$ resulting in a TLL, (b) $J_{\parallel}=-J_{\perp}$,
$V_{\perp}=V_{\parallel}=V_d=0$ not resulting in a TLL.}
\label{fig:LLnot}
\end{center}
\end{figure}

Despite the limitations of this RG method it has had some success in
predicting phases of various systems. Rather than attempt to obtain
the exact experimental values of $\alpha$ one could simply try to
replicate the line shape of $\alpha$ as the temperature decreases. In
Fig. \ref{fig:LLnot}(a) we show that the general line shape observed
in experiments is obtainable when $J_{\perp}$ is not particularly
large, although it must be {\em ferromagnetic} because we are still
bound by the condition $J_{\perp}<-\sfrac{4}{3}V_{\perp}$ if we wish
to have a TLL. In Fig. \ref{fig:LLnot}(b) we show that when we do not
have a TLL $\alpha$ may still adopt a variety of line shapes, some of
which strongly resemble a TLL as their turning point is very close to
$T=0$, in particular the $J_{\parallel}=-J_{\perp}=1/2$ case. Also,
by rescaling the interaction strengths it is possible to rescale
almost any $\alpha$ to have a very low temperature turning point.

If we rescale all interactions by the same factor $R$ so that
$(U,J_{n\parallel},J_{n\perp},V_{n\parallel},V_{n\perp})\rightarrow
(U,J_{n\parallel},J_{n\perp},V_{n\parallel},V_{n\perp})/R$ then,
because the initial couplings are linear in the interactions we can
define a new set of couplings $\tilde{g}_i$ in terms of these
rescaled interactions, $g_i(\ell=0)=R\tilde{g}_i(\ell=0)$. Rescaling
the RG flow equations in the same way gives $d \tilde{g}_{i}/
d\tilde{\ell} =\sum_{jk} A^{jk}_{i}\tilde{g}_{j}\tilde{g}_{k}$ where
$\tilde{\ell}=R\ell$ and because it is the ratios
$(J_{n\parallel},J_{n\perp},V_{n\parallel},V_{n\perp})/U$ which
essentially determine the phase both $g_i$ and $\tilde{g}_i$ should
flow towards the same solution and eventually describe the same
phase. However, this does not mean they will have the same scaling.
The rescaled temperature is $\tilde{T}=T_0e^{-\tilde{l}}$ and so
$\tilde{T}=T_0(T/T_0)^{-R}$ and therefore, by choosing an appropriate
$R$ we may rescale $\alpha$ so that its turning point is very close
to $T=0$ and the phase may closely resemble a TLL over a large
temperature scale. For example, the $J_{\parallel}=-J_{\perp}=4$
curve in Fig. \ref{fig:LLnot}(b) has a turning point at $T$=115 K,
but if we choose $R=2$ we rescale to $U=1$ and
$J_{\parallel}=-J_{\perp}=2$ which rescales the turning point of
$\alpha$ to $T=44$ K. Similarly, if we choose $R=4$ we obtain a
turning point at $T=6.4$ K.

\begin{figure}
\begin{center}
\includegraphics[width=8cm]{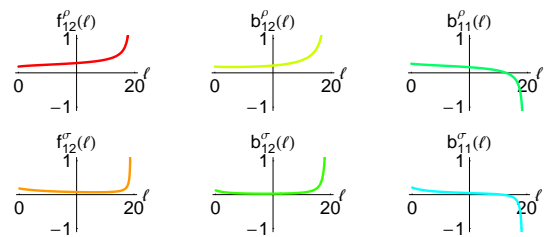}
\caption{(Color online) The renormalized coupling constants (rescaled by $4\pi v_q$)
at quarter-filling with $t=1$, $t_{\perp}=0.01$,
$U=2$, $J_{\parallel}=-J_{\perp}=0.5$ and $V_{\parallel,\perp,d}=0$.}
\label{fig:RGflow}
\end{center}
\end{figure}

In Fig (\ref{fig:RGflow}) we show the RG flow of the couplings for
the $U=2$, $J_{\parallel}=-J_{\perp}=1/2$ case. These couplings
mostly behave very much like one would expect in a TLL, with
$f^{\sigma}_{12}$, $b^{\sigma}_{12}$ and $b^{\sigma}_{11}$
approaching zero while $b^{\rho}_{11}$ and $f^{\rho}_{12}$ are fairly
constant, resulting in a fairly constant $\alpha$ over a large
temperature range. Only $b^{\rho}_{12}$ does not have typical TLL
behavior as it does not approach zero. Because of this the RG
eventually flows away from typical TLL behavior and the couplings
become large, in this case flowing towards a typical D-SC solution.
In the previous section it was mentioned that the region outlined in
red in Fig. \ref{fig:phaseLPB2-10}(a) is not a TLL, but may be
mistaken for one. This is because these couplings scale similarly to
the ones shown in Fig (\ref{fig:RGflow}), specifically
$b^{\rho}_{11}$ and $f^{\rho}_{12}$ (and therefore $\alpha$) remain
fairly constant over a significant $\ell$ range but at some point
they make a rapid change and approach values typical of a
superconductor. As the experimental data of LPB also hints at an
increase in the critical exponent $\alpha$ near $T=0$,~\cite{Wang06}
it cannot be ruled out that the observed scaling is indeed a close
crossover from TLL-like behavior to some superconducting or density
wave phase near zero temperature.

\section{Conclusions}

We have derived a Hamiltonian for a two-leg ladder which allows
consideration of generic short-range charge and spin interactions.
One can choose the interactions to extend only to nearest neighbors,
or one can choose to have interactions which extend across several
lattice sites. When increasing the range of the interactions the
number of variables inevitably increases. Rather than considering
each interaction strength as an independent variable and dealing with
all the associated problems, we can simply assume that the
interactions are inversely proportional to the distance between
lattice sites, thus keeping the number of parameters to a minimum.
Using this method we were able to solve the RG solutions for any
number of interactions while only needing three variables, $U$, $V$
and $J$, to describe the electron-electron interactions.

The Hamiltonian derived here is applicable to several different
materials, and not just those materials like LPB which have an
obvious ladder structure. Carbon nanotubes, for example, have an
hexagonal lattice structure which may be mapped onto a two-leg
ladder, and results obtained from two-leg ladder RG equations have
been applied to carbon nanotubes with nearest-neighbor
interactions.~\cite{Bunder08} However, carbon nanotubes are known to
support long-range interactions so the Hamiltonian presented here,
with slight modifications, would provided a more accurate picture of
the phases of a carbon nanotube.

Our RG analysis of LPB is somewhat limited because we have no
experimental data which gives any clear indication of the charge and
spin interaction strengths. It is important to note that this data
should not be obtained indirectly by attempting to fit the
experimental flow of the critical exponent $\alpha$ to numerical
solutions of $\alpha$ obtained from the RG equations as these
numerical solutions are not expected to be quantitatively accurate.
Because of these limitations we are unable to make a definite
statement concerning a TLL phase in LPB. The observed behavior may be
a true TLL phase, or it may simply be a non-TLL phase which strongly
resembles a TLL over a large temperature range. The power of these
one-loop RG solutions is that they are relatively simple and tend to
provide a qualitative description of the phase of the system. More
quantitative accuracy may possibly be achieved from second-loop or
higher order corrections.~\cite{Tsuchiizu06}

We acknowledge support from the National Science Council of Taiwan
through grants NSC-96-2112-M-007-004 and NSC-97-2112-M-007-022-MY3
and also support from the National Center for Theoretical Sciences in
Taiwan.

\end{document}